# Barriers to motion and rotation of graphene layers based on measurements of shear mode frequencies

Andrey M. Popov[a,1], Irina V. Lebedeva[b,c], Andrey A. Knizhnik[b,c], Yurii E. Lozovik[a,d], and Boris V. Potapkin[b,c]

[a]Institute of Spectroscopy of Russian Academy of Sciences, Fizicheskaya Street 5, Troitsk, Moscow Region, 142190, Russia,

[b]National Research Centre "Kurchatov Institute", Kurchatov Square 1, Moscow, 123182, Russia,

[c]Kintech Lab Ltd, Kurchatov Square 1, Moscow, 123182, Russia,

[d]Moscow Institute of Physics and Technology, Institutskii pereulok 9, Dolgoprudny, Moscow Region, 141701, Russia

**Abstract**

Both van der Waals corrected density functional theory and classical calculations show that the potential relief of interaction energy between layers of graphite and few-layer graphene can be described by a simple expression containing only the first Fourier components. Thus a set of physical quantities and phenomena associated with in-plane relative vibration, translational motion and rotation of graphene layers are interrelated and are determined by a single parameter characterizing the roughness of the potential energy relief. This relationship is used to estimate the barriers to relative motion and rotation of graphene layers based on experimental measurements of shear mode frequencies.

[1]Corresponding author. Tel.: +7 496 7510881. Fax: +7 496 7510886.

*E-mail addresses:* popov-isan@mail.ru (A.M. Popov), lebedeva@kintechlab.com (I.V. Lebedeva), knizhnik@kintechlab.com (A.A. Knizhnik), lozovik@isan.troitsk.ru (Yu.E. Lozovik

1. **Introduction**

Due to unique electrical, mechanical and chemical properties, new two-dimensional carbon nanostructure, graphene [1], holds great promise for a variety of applications. A wide set of properties and applications of few-layer graphene are related to the van der Waals interaction between graphene layers. The possibility for graphene layers to form incommensurate configurations upon their relative rotation was shown to be responsible for ultra low static friction [2–4] and fast thermally activated diffusion [5, 6] of a graphene flake on a graphene layer. The way to control the diffusion coefficient of a graphene flake on a graphene layer using external fields fixing a commensurate or incommensurate orientation of the flake was proposed [6]. Self-retracting motion of graphite microflakes, i.e. retraction of graphite flakes back into the graphite stacks on their extension arising from the van der Waals interaction, was demonstrated experimentally [7]. The theoretical analysis [8] revealed that transition of graphene flakes to incommensurate configurations with low static friction is also crucial for their self-retracting motion. Thus all the phenomena listed above are determined by value of the barrier to rotation of the graphene flake from commensurate to incommensurate states. Ultrafast damping of small relative vibrations and absence of telescopic oscillations of graphene layers makes them suitable for the use in fast-responding nanorelays and memory cells based on relative motion of graphene layers [8–10]. Mechanical properties of bilayer graphene, such as the critical elongation of one of the layers at which incommensurability defects start to form and the threshold force required to start relative motion of graphene layers, were also shown to be determined by the barrier to relative motion of graphene layers [11]. Nanoresonators based on flexural vibrations of suspended graphene were implemented [12]. The calculated frequency of flexural vibrations of few-layer graphene was found to be sensitive to the interlayer shear rigidity [13]. The interlayer interaction can be responsible for the drastic increase of thermal conductivity of graphene with decreasing the number of layers [14] and is also important for operation of graphene-based touch-screen panel devices [15] and mechanical properties of graphene-based composites [16]. Therefore adequate description of these and other phenomena and nanoelectromechanical systems requires knowledge of quantitative characteristics of the interaction between graphene layers, such as barriers to their relative motion and rotation.

The weakness of van der Waals forces binding graphene layers leads to experimental difficulties when measuring physical quantities characterizing the interaction between graphene layers. The data on the interlayer binding energy show significant scatter depending on the experimental approach [17 – 19]. No experimental data on the corrugation of the potential relief of interlayer interaction energy are available until now. The value of the critical shear strength for graphite measured in the only known experiment [20] is related to macroscopic structural defects of the graphite sample. The experiments with the friction force microscope allow to investigate only a small region of the potential relief of interlayer interaction energy due to slip-stick motion of the graphene flake attached to the microscope tip on the graphite surface [3, 4].

On the other hand, first-principles theoretical methods, such as density functional theory (DFT), also meet serious difficulties when applied to interacting graphene layers [21 – 23]. Despite the recent progress in incorporation of the van der Waals interaction in DFT calculations, accurate quantitative description of the potential relief of interaction energy of graphene layers still remains a challenge. Barriers to relative motion of graphene layers and magnitudes of corrugation of the potential energy relief obtained by DFT calculations with account of the van der Waals interaction [9, 10, 24] or by standard DFT calculations [5, 6, 25, 26] (for the equilibrium interlayer distance in the case of the local density approximation [5, 6] and for a pre-determined interlayer distance in the case of the generalized gradient approximation [25, 26]) range from 1 to 2 meV/atom and from 10 to 20 meV/atom, respectively.

However we show here that the potential relief of interaction energy between layers of graphite and few-layer graphene can be described by a simple expression containing only the first Fourier components with an error that is an order of magnitude less than the mentioned above discrepancy in the barriers calculated using different first-principles methods. Such a simple shape of the potential relief means that a set of physical quantities which are determined by this shape are described by the same single parameter characterizing the roughness of the potential energy relief, i.e. these quantities are interrelated. Particularly, the interrelated physical quantities include the barrier to in-plane relative translational motion of graphene layers, the barrier to in-plane relative rotation of graphene layers from commensurate to incommensurate states, the magnitude of corrugation of the potential relief of interlayer interaction energy and the frequency of in-plane relative vibrations of graphene layers. As opposed to such characteristics of the potential relief of interlayer interaction energy as the magnitude of corrugation and the barriers to relative motion and rotation of the layers, the phonon spectrum of graphite was measured experimentally [27, 28]. Lately Raman measurements of the shear mode in few-layer graphene were also reported [29]. Thus we suggest that the barriers to relative motion and rotation of graphene layers as well as the magnitude of corrugation of the potential energy relief can be estimated on the basis of frequency measurements.

2. **Results and discussion**

A simple expression containing only the first Fourier components can be suggested for the interaction energy of two graphene layers [5, 6, 9]

$$U(x,y,z) = U_1(z)\left(1.5 + cos\left(2k_1 x - \frac{2\pi}{3}\right) - 2cos\left(k_1 x - \frac{\pi}{3}\right)cos(k_2 y)\right) + U_0(z), \quad (1)$$

where $k_1 = 2\pi/(3l)$, $k_2 = 2\pi/(\sqrt{3}l)$, $l = 1.42$ Å is the bond length in graphene, $x$ and $y$ axes are chosen along the armchair and zigzag directions, respectively. It is surprising that this simple expression which is based only on the symmetry considerations, as was recently shown, fits closely the results of calculations of interlayer interaction energy in bilayer graphene both in the framework of the density functional theory (DFT) and using classical potentials [5, 6, 9] (see Fig. 1). In the present Letter, we demonstrate that Eq. (1) is also adequate for description of the interlayer interaction in trilayer graphene and graphite. Note that such a simple shape of the potential energy relief is not an exclusive property of the interaction between graphene layers. For example, the first Fourier components are also sufficient to describe the potential reliefs of interwall interaction energy of double-walled nanotubes with commensurate nonchiral walls within accuracy less than 1% in the case of the use of the Lennard-Jones potential [30] and within the accuracy of calculations in the case of the use of the density functional methods [31 – 33].

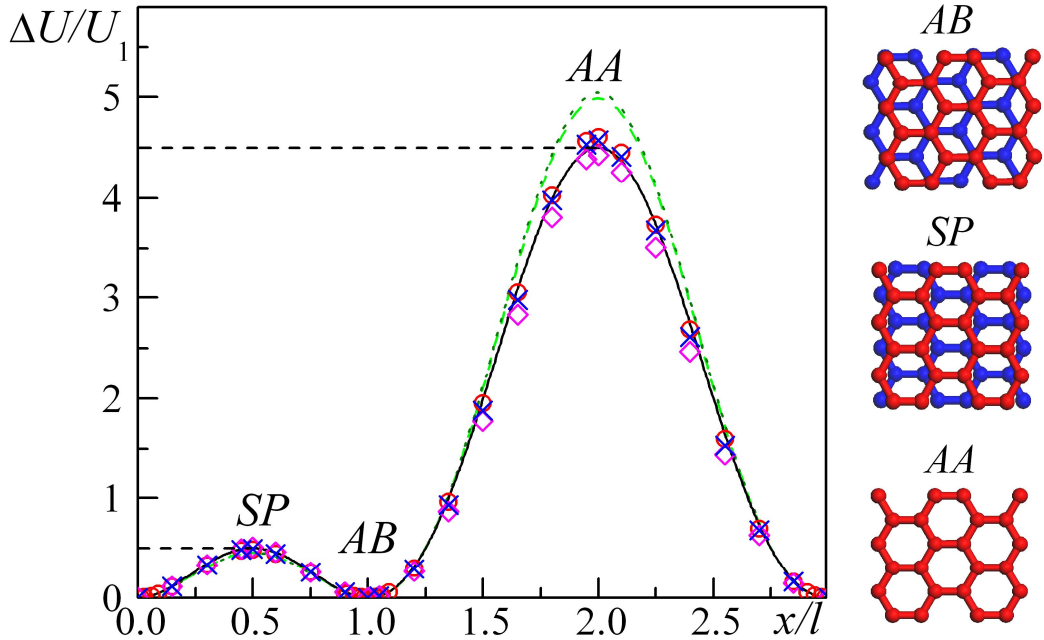

**Figure 1**. (Left) Interaction energy $\Delta U / U_1 = (U - U_0)/U_1$ between commensurate graphene layers as a function of the relative displacement $x/l$ of the layers in the armchair direction corresponding to approximation (1) (black solid line). The results of DFT-D calculations for bilayer (red circles), trilayer graphene (blue crosses) and graphite (magenta diamonds) and the results of calculations using the Lennard-Jones potential for bilayer graphene (light green dashed line) and graphite (dark green dotted line) are shown. The corresponding values of $U_1$ and $U_0$

are listed in Table 1. The results of calculations for bilayer graphene are taken from paper [9]. The black dashed lines show the values of the barrier $\Delta E_{\mathrm{SP}}/U_1 = 1/2$ to relative motion of graphene layers and the magnitude $\Delta E_{\mathrm{AA}}/U_1 = 9/2$ of corrugation of the potential energy relief following from approximation (1). (Right) Structures corresponding to the AB, SP and AA stackings of commensurate graphene layers.

To obtain the potential reliefs of interlayer interaction energy in trilayer graphene and graphite at the ABA stacking we have performed dispersion-corrected density functional theory (DFT-D) calculations using the VASP code [34] with the generalized gradient approximation (GGA) density functional of Perdew, Burke, and Ernzerhof [35] corrected with the dispersion term [36]. The periodic boundary conditions are applied to a 4.27 Å x 2.47 Å x 20 Å model cell for trilayer graphene and a 4.27 Å x 2.47 Å x 6.44 Å model cell containing two graphene layers for graphite. The basis set consists of plane waves with the maximum kinetic energy of 500 eV. The interaction of valence electrons with atomic cores is described using the projector augmented-wave method (PAW) [37]. Integration over the Brillouin zone is performed using the Monkhorst-Pack method [38] with 24x36x1 k-point sampling for trilayer graphene and 24x36x16 k-point sampling for graphite. The graphene layers are placed at the equilibrium interlayer distance of 3.25 Å for trilayer graphene and 3.22 Å for graphite. In the calculations of the potential energy relief for trilayer graphene, the middle graphene layer is rigidly shifted parallel to the others. In the calculations for graphite, one of the graphene layers in the model cell is rigidly shifted parallel to the other layer. Account of structure deformation induced by the interlayer interaction was shown to be inessential for the shape of the potential relief of interaction energy between graphene-like layers, such as the interwall interaction of carbon nanotubes [30] and the intershell interaction of carbon nanoparticles [39, 40]. The calculated dependences of interlayer interaction energy in trilayer graphene and graphite on the relative displacement of the layers in the armchair direction are shown in Fig. 1.

We have also performed calculations of the potential reliefs of interlayer interaction energy in trilayer graphene and graphite using the Lennard-Jones potential

$$U_{LJ} = 4\varepsilon\left(\left(\frac{\sigma}{r}\right)^{12} - \left(\frac{\sigma}{r}\right)^{6}\right) \qquad (2)$$

The parameters of the Lennard–Jones potential $\varepsilon = 2.76$ meV, $\sigma = 3.39$ Å were previously fitted by us to reproduce the interlayer binding energy, interlayer distance and $c$-axis compressibility of graphite [9]. The cutoff distance of the potential is equal to 25 Å. The dependences of interlayer interaction energy in trilayer graphene and graphite on the relative displacement of the layers in the armchair direction calculated using the Lennard-Jones potential at the equilibrium interlayer distance are also shown in Fig. 1.

It is seen from Fig. 1 that the potential reliefs of interlayer interaction energy in few-layer graphene and graphite can be closely reproduced using approximation (1). The parameters $U_0$ and $U_1$ of approximation (1) fitted to the results of DFT-D and classical calculations at equilibrium interlayer distances $d_{eq}$ are given in Table 1. The relative root-mean-square deviations $\delta U / U_1$ of approximation (1) from the calculated potential reliefs of interlayer interaction energy are within 0.13 for the results of DFT-D calculations and within 0.23 for the results of calculations using the Lennard-Jones potential. Some differences in the parameters $U_0$ and $U_1$ for bilayer, trilayer graphene and graphite are explained by the interaction of non-adjacent graphene layers, which is not considered in Eq. (1). The relative root-mean-square deviation $\delta U / U_1$ is the smallest for bilayer graphene.

**Table 1**. Parameters $U_0$, $U_1$ of approximation (1) per atom of one of graphene layers and relative root-mean-square deviations $\delta U / U_1$ of approximation (1) from the potential reliefs of interaction energy between graphene layers obtained by the DFT-D calculations and using the Lennard-Jones potential for *n*-layer graphene and graphite ($n \to \infty$) at the equilibrium interlayer distance $d_{eq}$.

| Method | $n$ | $d_{eq}$ (Å) | $U_0$ (meV/atom) | $U_1$ (meV/atom) | $\delta U / U_1$ |
|---|---|---|---|---|---|
| DFT-D | 2 | 3.25 | −50.59 | 4.24 | 0.0428 |
| | 3 | 3.25 | −52.68 | 4.14 | 0.0591 |
| | ∞ | 3.22 | −57.04 | 4.65 | 0.128 |
| Lennard- | 2 | 3.384 | −45.67 | 0.178 | 0.208 |
| Jones | 3 | 3.366 | −48.06 | 0.194 | 0.218 |
| potential | ∞ | 3.340 | −51.89 | 0.222 | 0.223 |

It follows from approximation (1) that all characteristics of the potential relief of interaction energy of graphene layers can be expressed through a single parameter $U_1$. The global energy minimum $U_0$ for the interaction of two graphene layers is reached at the AB stacking ($x = 0, y = 0$; see Fig. 1). The relative energy of the saddle-point (SP) stacking $\Delta E_{SP} = U(x = l/2, y = 0) - U_0 = 0.5 U_1$ (see Fig. 1) corresponds to the barrier to relative motion of the layers. The relative energy of the AA stacking $\Delta E_{AA} = U(x = 2l, y = 0) - U_0 = 4.5 U_1$ (see Fig. 1) determines to the magnitude of corrugation of the potential relief of interlayer interaction energy. Upon relative rotation of graphene layers to incommensurate states, the potential relief of interaction energy between the layers becomes smooth [2 – 6, 8]. The interaction energy of graphene layers in such states can be estimated as an average of the interaction energy (1) over in-plane relative displacements of the layers in the commensurate state $U_{in} = \langle U \rangle_{x,y}$. Thus the barrier to relative rotation of graphene layers from the global energy minimum can be found as $\Delta E_{in} = U_{in} - U_0 = 1.5 U_1$.

Based on Eq. (1), the frequency of the shear mode in $n$-layer graphene and graphite ($n \to \infty$) can be also expressed through the single parameter $U_1$

$$f_0 = \frac{1}{2\pi}\sqrt{\frac{(n-1)}{\mu}\frac{\partial^2 U}{\partial x^2}\bigg|_{x=0}} = \frac{1}{l}\sqrt{\frac{(n-1)U_1}{3\mu}}, \qquad (3)$$

where $\mu = pm/2$ for $n = 2p$ and $\mu = p(p+1)m/(2p+1)$ for $n = 2p+1$, $m$ is the mass of a carbon atom and $p$ is an integer. Therefore based on the experimental data on the shear mode frequencies $f_0$, the parameters $U_1$ for few-layer graphene and graphite can be found as $U_1 = 3\mu(lf_0)^2/(n-1)$. The above expressions for $\Delta E_{SP}$, $\Delta E_{AA}$ and $\Delta E_{in}$ in terms of $U_1$ can be used to estimate these characteristics of the potential relief of interlayer interaction energy (see Table 2).

From Table 2, it follows that the barriers $\Delta E_{SP}$ and $\Delta E_{in}$ to relative motion and rotation of two graphene layers and the magnitude $\Delta E_{AA}$ of corrugation of the potential relief of interaction energy between two graphene layers are $\Delta E_{SP} \sim 1.7$ meV/atom, $\Delta E_{in} \sim 5$ meV/atom and the $\Delta E_{AA} \sim 15$ meV/atom, respectively. These values of $\Delta E_{SP}$, $\Delta E_{in}$ and $\Delta E_{AA}$ can be used to revise theoretical description of the phenomena related to interaction of graphene layers. As an example, let us consider the commensurate-incommensurate phase transition in bilayer graphene which is observed upon stretching or compression of one of the layers. Based on the expressions given in paper [11] and assuming $\Delta E_{SP} = 1.7$ meV/atom, we estimate the critical unit elongation of one of the graphene layers in the armchair direction at which the first incommensurability defect forms to be $\delta = 3.6 \cdot 10^{-3}$. The threshold force required to start relative motion of graphene layers in the armchair direction is found to be $F_0 = 0.19$ nN/Å (per unit width of the layers in the direction perpendicular to the elongation). It should also be noted that the values of $\Delta E_{SP}$, $\Delta E_{in}$ and $\Delta E_{AA}$ estimated above can be reproduced by the recently developed classical potential [9] if the parameter $C$ of this potential is set at 29.5 meV.

**Table 2**. Barriers $\Delta E_{SP}$ and $\Delta E_{in}$ to relative motion and rotation of two graphene layers, magnitudes $\Delta E_{AA}$ of corrugation of the potential relief of interaction energy between the graphene layers and parameters $U_1$ for approximation (1) of the potential energy relief per atom of one of the layers estimated on the basis of experimental data on frequencies $f_0$ of the shear mode in $n$-layer graphene and graphite ($n \to \infty$).

| Ref. | $n$ | $f_0$ (cm$^{-1}$) | $U_1$ (meV/atom) | $\Delta E_{SP}$ (meV/atom) | $\Delta E_{in}$ (meV/atom) | $\Delta E_{AA}$ (meV/atom) |
|---|---|---|---|---|---|---|
| 29 | 2 | 32 | 3.43 | 1.71 | 5.14 | 15.4 |
|  | 3 | 38 | 3.22 | 1.61 | 4.84 | 14.5 |
|  | 4 | 41 | 3.75 | 1.88 | 5.63 | 16.9 |
|  | 5 | 42 | 3.55 | 1.77 | 5.32 | 16.0 |
|  | ∞ | 44 | 3.24 | 1.62 | 4.86 | 14.6 |
| 27 | ∞ | 45 | 3.39 | 1.69 | 5.09 | 15.3 |
| 28 | ∞ | 42 | 2.96 | 1.48 | 4.43 | 13.3 |

To check that the presence of some defects in experimental samples should not strongly influence the potential relief of interaction energy between graphene layers we have also performed spin-unrestricted DFT-D calculations for bilayer graphene with vacancies. In these calculations, the size of the model cell is 21.4 Å x 22.2 Å x 14 Å. Integration over the Brillouin zone is performed using 5 x 5 x 1 k-point sampling. A single reconstructed vacancy is introduced into one of the layers, which corresponds to the density of defects 0.21 $nm^{-2}$. The structures of the layers with and without the vacancy are separately relaxed. Then the energies of configurations corresponding to the AB, SP and AA stackings are calculated at the equilibrium interlayer distance of 3.25 Å. It is found that in this system adjacent energy minima represented by the AB stacking slightly differ in energy (by 0.27 meV/atom). The calculated relative energies $\Delta E_{SP}$ and $\Delta E_{AA}$ of the SP and AA stackings associated with the barrier to relative motion of the graphene layers and the magnitude of corrugation of the potential energy relief deviate from the values for perfect bilayer graphene by less than 11% and 6%, respectively. In high-quality few-layer graphene or graphite the density of defects is less than 15 $\mu m^{-2}$ [41], i.e. $10^4$ times smaller than in the system under consideration. Thus the disturbance of the potential relief of interlayer interaction energy induced by point defects in experimental samples should be negligibly small. The same conclusion was made in our previous publication for vacancy and Stone–Wales defects on the basis of calculations using the Lennard-Jones potential [6].

It should also be noted that in real few-layer graphene not only defects but also ripples [42, 43] can influence the potential relief of interlayer interaction energy. The ripples strongly decrease in height with increasing the number of graphene layers [42 – 44] so this effect should the most prominent for bilayer graphene. The experimental studies [43] show that in bilayer graphene the ripples provide the deviation of the surface normal from its mean direction by the angle $\varphi = 2^\circ$. To estimate the effect of such ripples on the potential energy relief of bilayer graphene we have performed DFT-D calculations for the system in which the graphene layers experience coherent sinusoidal out-of-plane waves $\Delta z = A \sin(2\pi x / L)$, where the $x$ axis is directed along the armchair direction. The 51.3 Å x 2.47 Å x 14 Å model cell is considered. Integration over the Brillouin zone is performed using 3 x 36 x 1 k-point sampling. The period of the wave $L = 51.3$ Å equals the size of the model cell along the armchair direction. The amplitude of the wave $A = 0.3$ Å is chosen so that the ratio of the amplitude $A$ to the period $L$ corresponds to the experimental value $2\pi A / L \approx \varphi = 0.035$. The calculations performed show that the waviness leads to the decrease of the relative energies $\Delta E_{SP}$ and $\Delta E_{AA}$ of the SP and AA stackings by 7% and 1%, respectively. Some decrease of the barrier to relative motion of graphene layers and the magnitude of corrugation of the potential energy relief because of the presence of ripples should also take place in graphene with three and more layers. In graphite ripples are completely absent [43]. Therefore the parameters $U_1$ of the potential relief of interlayer interaction energy should be slightly different for few-layer graphene and graphite. However, the scatter in the parameters $U_1$ (see Table 2) estimated on the basis of the experimental measurements of shear mode frequencies for few-layer graphene and graphite shows that the accuracy of such measurements is insufficient to reveal the influence of ripples on the potential relief of interaction energy of graphene layers.

3. **Conclusions**

It has been shown that the potential relief of interaction energy between graphene layers can be described using the simple expression containing only the first Fourier components. That is all physical properties related to in-plane relative displacement of the layers are determined by the single parameter. This observation is used to reproduce the potential relief of interaction energy between graphene layers on the basis of the experimental data on the shear mode in few-layer graphene and graphite. The barriers $\Delta E_{SP}$ and $\Delta E_{in}$ to relative motion and rotation of two graphene layers and the magnitude $\Delta E_{AA}$ of corrugation of the potential relief of interaction energy between two graphene layers are estimated to be $\Delta E_{SP} \sim 1.7$ meV/atom, $\Delta E_{in} \sim 5$ meV/atom and the $\Delta E_{AA} \sim 15$ meV/atom, respectively.

**Acknowledgements**

This work has been supported by the RFBR grants 11-02-00604 and 12-02-90041-Bel. The calculations are performed on the SKIF MSU Chebyshev supercomputer, the MVS-100K supercomputer at the Joint Supercomputer Center of the Russian Academy of Sciences and the Multipurpose Computing Complex NRC "Kurchatov Institute".

**References**

[1] K.S. Novoselov, A.K. Geim, S.V. Morozov, D. Jiang, Y. Zhang, S.V. Dubonos, I.V. Grigorieva and A.A. Firsov, Science 306 (2004) 666.

[2] M. Dienwiebel, G.S. Verhoeven, N. Pradeep, J.W.M. Frenken, J.A. Heimberg, H.W. Zandbergen, Phys. Rev. Lett. 92 (2004) 126101.

[3] G.S. Verhoeven, M. Dienwiebel, J.W.M. Frenken, Phys. Rev. B 70 (2004) 165418.

[4] N. Sasaki, K. Kobayashi, M. Tsukada, Phys. Rev. B 54 (1996) 2138.

[5] I.V. Lebedeva, A.A. Knizhnik, A.M. Popov, O.V. Ershova, Yu.E. Lozovik, B.V. Potapkin, Phys. Rev. B 82 (2010) 155460.

[6] I.V. Lebedeva, A.A. Knizhnik, A.M. Popov, O.V. Ershova, Yu.E. Lozovik, B.V. Potapkin, J. Chem. Phys. 134 (2011) 104505.

[7] Q. Zheng, B. Jiang, S. Liu, Y. Weng, L. Lu, Q. Xue, J. Zhu, Q. Jiang, S. Wang, L. Peng, Phys. Rev. Lett. 100 (2008) 067205.

[8] A.M. Popov, I.V. Lebedeva, A.A. Knizhnik, Yu.E. Lozovik, B.V. Potapkin, Phys. Rev. B 84 (2011) 245437.

[9] I.V. Lebedeva, A.A. Knizhnik, A.M. Popov, Yu.E. Lozovik, B.V. Potapkin, Phys. Chem. Chem. Phys. 13 (2011) 5687.

[10] I.V. Lebedeva, A.A. Knizhnik, A.M. Popov, Yu.E. Lozovik, B.V. Potapkin, Physica E (2011) DOI:10.1016/j.physe.2011.07.018 (in print).

[11] A.M. Popov, I.V. Lebedeva, A.A. Knizhnik, Yu.E. Lozovik, B.V. Potapkin, Phys. Rev. B 84 (2011) 045404.

[12] J.S. Bunch, A.M. van der Zande, S.S. Verbridge, I.W. Frank, D.M. Tanenbaum, J.M. Parpia, H.G. Craighead, P.L. McEuen, Science 315 (2007) 490.

[13] Y. Liu, Z. Xu, Q. Zheng, Journal of the Mechanics and Physics of Solids 59 (2011) 1613.

[14] S. Ghosh, W. Bao, D.L. Nika, S. Subrina, E.P. Pokatilov, C.N. Lau, A.A. Balandin, Nature Materials 9 (2010) 555.

[15] S. Bae, H. Kim, Y. Lee, X. Xu, J.-S. Park, Y. Zheng, J. Balakrishnan, T. Lei, H.R. Kim, Y.I. Song, Y.-J. Kim, K.S. Kim, B. Özyilmaz, J.-H. Ahn, B.H. Hong, S. Iijima, Nature Nanotechnology 5 (2010) 574.

[16] D.A. Dikin, S. Stankovich, E.J. Zimney, R.D. Piner, G.H.B. Dommett, G. Evmenenko, S.B.T. Nguyen, R.S. Ruoff, Nature 448 (2007) 457.

[17] L.X. Benedict, N.G. Chopra, M.L. Cohen, A. Zettl, S.G. Louie, V.H. Crespi, Chem. Phys. Lett. 286 (1998) 490.

[18] L.A. Girifalco, R.A. Lad, J. Chem. Phys. 25 (1956) 693.

[19] R. Zacharia, H. Ulbricht, T. Hertel, Phys. Rev. B 69 (2004) 155406.

[20] D.E. Soule, C.W. Nezbeda, J. Appl. Phys. 39 (1968) 5122.

[21] L.A. Girifalco, M. Hodak, Phys. Rev. B 65 (2002) 125404.

[22] H. Rydberg, M. Dion, N. Jacobson, E. Schröder, P. Hyldgaard, S.I. Simak, D.C. Langreth, B.I. Lundqvist, Phys. Rev. Lett. 91 (2003) 126402.

[23] M. Hasegawa, K. Nishidate, Phys. Rev. B 70 (2004) 205431.

[24] O.V. Ershova, T.C. Lillestolen, E. Bichoutskaia, Phys. Chem. Chem. Phys. 12 (2010) 6483.

[25] A.N. Kolmogorov, V.H. Crespi, Phys. Rev. Lett. 85 (2000) 4727.

[26] A.N. Kolmogorov, V.H. Crespi, Phys. Rev. B 71 (2005) 235415.

[27] R. Nicklow, N. Wakabayashi, H.G. Smith, Phys. Rev. B 5 (1972) 4951.

[28] P.C. Eklund, J.M. Holden, R.A. Jishi, Carbon 33 (1995) 959.

[29] P.H. Tan, W.P. Han, W.J. Zhao, Z.H. Wu, K. Chang, H. Wang, Y.F. Wang, N. Bonini, N. Marzari, N. Pugno, G. Savini, A. Lombardo, A.C. Ferrari, Nature Mat. 11 (2012) 294.

[30] A.V. Belikov, Yu.E. Lozovik, A.G. Nikolaev, A.M. Popov, Chem. Phys. Lett. 385 (2004) 72.

[31] E. Bichoutskaia, A.M. Popov, A. El-Barbary, M.I. Heggie, Yu.E. Lozovik, Phys. Rev. B 71 (2005) 113403.

[32] A.M. Popov, Yu.E. Lozovik, A.S. Sobennikov, A.A. Knizhnik, JETP 108 (2009) 621.

[33] E. Bichoutskaia, A.M. Popov, Yu.E. Lozovik, O.V. Ershova, I.V. Lebedeva, A.A. Knizhnik, Phys. Rev. B 80 (2009) 165427.

[34] G. Kresse, J. Furthmüller, Phys. Rev. B 54 (1996) 11169.

[35] J.P. Perdew, K. Burke, M. Ernzerhof, Phys. Rev. Lett. 77 (1996) 3865.

[36] S. Grimme, J. Comp. Chem. 27 (2006) 1787.

[37] G. Kresse, D. Joubert, Phys. Rev. B 59 (1999) 1758.

[38] H.J. Monkhorst, J.D. Pack, Phys. Rev. B 13 (1976) 5188.

[39] Yu. E. Lozovik, A. M. Popov, Chem. Phys. Lett. 328 (2000) 355.

[40] Yu. E. Lozovik, A. M. Popov, Phys. Solid State 44 (2002) 186.

[41] S. Wu, R. Yang, D. Shi, G. Zhang, Nanoscale 4 (2012) 2005.

[42] J. C. Meyer, A. K. Geim, M. I. Katsnelson, K. S. Novoselov, D. Obergfell, S. Roth, C. Girita, A. Zettl, Solid State Comm.143 (2007) 101.

[43] J. C. Meyer, A. K. Geim, M. I. Katsnelson, K. S. Novoselov, T. J. Booth and S. Roth, Nature 446 (2007) 60.

[44] K. V. Zakharchenko, J. H. Los, M. I. Katsnelson, A. Fasolino, Phys. Rev. B 81 (2010) 235439.